\documentclass[12pt,12pt,sort&compress]{article}
\usepackage[T1]{fontenc}
\usepackage[latin9]{inputenc}
\usepackage[active]{srcltx}
\usepackage{float}
\usepackage{amsmath}
\usepackage{amssymb}
\usepackage{graphicx}
\usepackage[numbers]{natbib}

\makeatletter
\newcommand{\lyxaddress}[1]{
	\par {\raggedright #1
	\vspace{1.4em}
	\noindent\par}
}

\@ifundefined{date}{}{\date{}}

\usepackage{indentfirst}
\usepackage{amsfonts}
\usepackage[T1]{fontenc}
\usepackage{ae,aecompl}
\usepackage{sidecap}
\usepackage[section]{placeins}
\usepackage{epsf}

\usepackage{fancyhdr}

\makeatother

\begin{document}
\title{The origin of strings and rings in the atomic dynamics of disordered
systems}
\author{Omar Hussein, Yang Li, and Yuri Mishin}
\maketitle

\lyxaddress{Department of Physics and Astronomy, MSN 3F3, George Mason University,
Fairfax, Virginia 22030, USA}
\begin{abstract}
\noindent It has long been believed that the atomic dynamics in disordered
structures, such as undercooled liquids and pre-melted interfaces,
are characterized by collective atomic rearrangements in the form
of quasi-one-dimensional chains of atomic displacements (strings)
and their closed forms (rings). Here, we show by molecular dynamics
(MD) simulations that strings involving more than a few atoms do not
form by a single collective event. Instead, they represent trajectories
of propagating local density perturbations, which we call densitons.
The atoms on this trajectory are almost indistinguishable from their
environments except for the moving head of the string (densiton).
A densiton migrates by either single-atom jumps or a concerted rearrangement
of 2-3 atoms. The simulations reveal a remarkable similarity between
the strings in disordered and crystalline structures, in which the
densitons localize into point defects. This work calls for a significant
reinterpretation of the string concept and instead proposes a densiton
model of the atomic dynamics.
\end{abstract}

\section{Introduction\label{sec:Introduction}}

The atomic dynamics in fully or partially disordered materials controls
many of their properties, such as diffusive mass transport, viscosity,
and heat conduction \citep{Berthier2011,Derlet:2018aa,Derlet:2020aa,derlet2021viscosity,Derlet:2021aa,AlSm1}.
Despite decades of dedicated research efforts, the atomic dynamics
of disordered systems remains one of the least understood phenomena
in materials physics. One of the prominent features of such systems
is the dynamic heterogeneity in space and time \citep{Donati_1999,Douglas98,Kob_97,AlSm1,chandler2010dynamics,jung2005dynamical,10.1063/1.1644539,Karmakar:2009aa,Cui:2001aa}.
Namely, the atomic mobility is distributed highly non-uniformly across
the system, and this distribution constantly changes over time.

The dynamic heterogeneity has been investigated by several experimental
probes \citep{Marcus:1999aa,Cui:2001aa}. However, the most detailed
information about the underlying atomic mechanisms comes from computer
simulations. In particular, molecular dynamics (MD) simulations have
revealed that atomic displacements often occur in the form of chains,
or strings, encompassing a group of atoms whose displacements are
such that one atom jumps into the previous location of another \citep{Donati_1999,Douglas98,10.1063/1.1644539,Kob_97}.
That other atom, in turn, jumps into the previous position of yet
another atom, and so on. Occasionally, a string can be closed to itself,
forming what is called a ring. Strings and rings were observed in
amorphous solids and undercooled glass-forming liquids \citep{Donati_1999,Douglas98,Kob_97,AlSm1,chandler2010dynamics,jung2005dynamical,10.1063/1.1644539},
polymers \citep{AlSm1}, super-ionic solids \citep{annamareddy_superionic},
disordered grain boundaries \citep{Warren2009,Zhang2006,Chesser:2022aa,Mishin:2015ac},
phase boundaries \citep{Al-Si-IPB,Chesser:2024aa}, dislocation cores
\citep{Chesser:2024aa}, and open surfaces \citep{Zhang2010,Salez:2015aa}.
It is generally believed that a string is a quasi-one-dimensional
dynamic object representing a collective (also called cooperative
or correlated) displacement of a group of atoms \citep{Erhart2020,10.1063/1.1644539,Donati_1999,Douglas98,Kob_97,AlSm1}.

\begin{figure}
\includegraphics[width=1\textwidth]{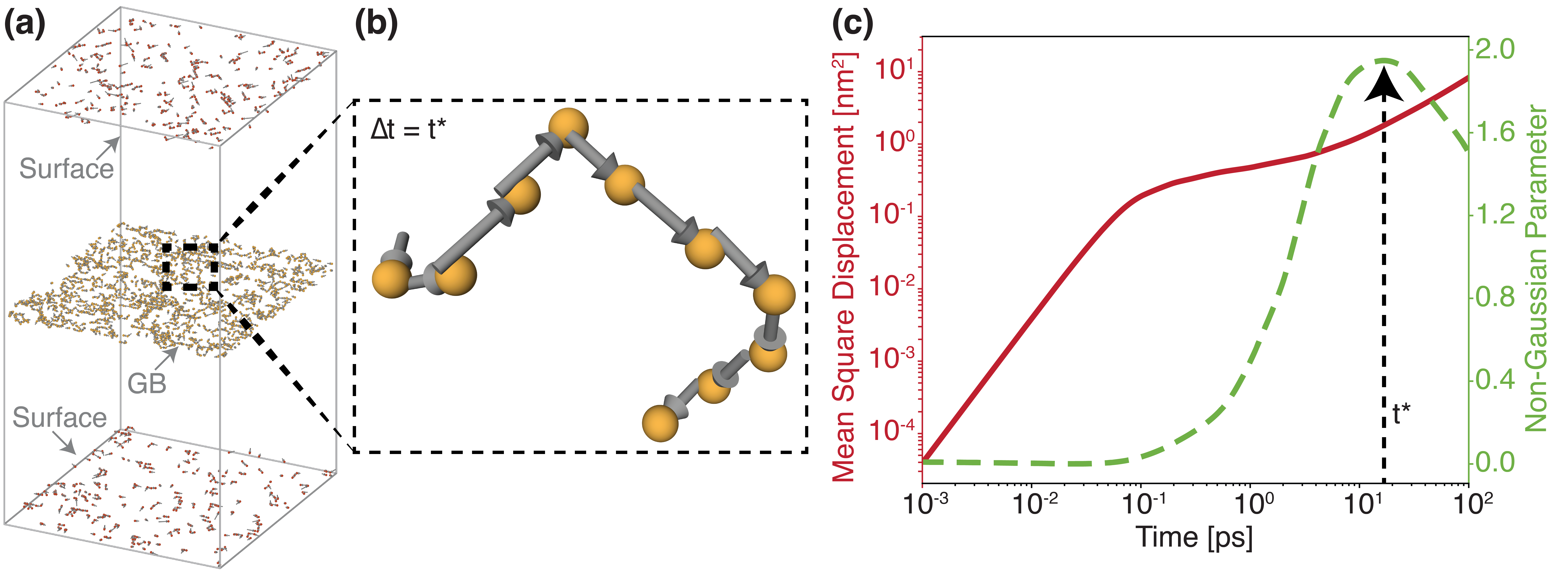}

\caption{The traditional method of string identification in MD simulations.
(a) A bicrystal of copper with a $\Sigma29$ $(250)$ tilt grain boundary
at 1000 K. The snapshot visualizes only mobile atoms with displacements
larger than $1.6r_{0}$ during a time interval of $\Delta t=t^{*}$.
The atoms are colored according to their location, with yellow and
red representing the grain boundary and surface atoms, respectively.
(b) Snapshot of a typical string formed at the grain boundary. The
gray arrows represent atomic displacements during the time $\Delta t$.
The string was identified by the methods traditionally applied to
glass-forming fluids. (c) Mean square displacement $\left\langle r^{2}\right\rangle $
and the non-Gaussian parameter (NGP) of the system as a function of
$\Delta t$. The time $t^{*}=17$ ps corresponds to the maximum of
the NGP. \label{fig:1}}
\end{figure}

In MD simulations, the strings are revealed by the following procedure
\citep{Douglas98}. First, mobile atoms are identified by the criterion
that their displacements $\Delta r$ during a chosen time interval
$\Delta t$ lie within the range $0.4r_{0}<\Delta r<1.2r_{0}$, where
$r_{0}$ is an average interatomic distance in the system. Here, the
lower bound eliminates immobile atoms and the upper bound eliminates
multiple atomic hops. The second criterion selects pairs $(i,j)$
of mobile atoms that remain nearest neighbors of each other during
the interval $\Delta t$: $\text{min}(|\mathbf{r}_{i}(t)-\mathbf{r}_{j}(0)|,|\mathbf{r}_{j}(t)-\mathbf{r}_{i}(0)|)<0.43r_{0}$.
Mobile pairs satisfying this criterion and sharing an atom are then
joined together into a mobile cluster, which often looks like a string
or a ring. (The hyper-parameters in the above criteria are slightly
adjusted for each particular system.) The string size measured by
this procedure depends on the chosen time interval $\Delta t$. In
most cases, $\Delta t$ is set as the position $t^{*}$ of the maximum
of the non-Gaussian parameter $\mathrm{NGP}=3\left\langle (\Delta r)^{4}\right\rangle /5\left\langle (\Delta r)^{2}\right\rangle ^{2}-1$
as a function of time. Alternatively, $\Delta t$ is identified with
the maximum of the average string length computed as a function of
time, which is usually close to $t^{*}$ \citep{10.1063/1.1644539}.
Yet another method identifies $\Delta t$ with the inflection point
on the mean-square displacement (MSD) plot as a function of time,
giving a similar value. Typical string length identified by the above
criteria varies from $n=2$ to $3$ atoms at high temperatures to
$n=10$ or more atoms at low temperatures. As an example, Fig.~\ref{fig:1}
demonstrates a 10-atom string in a disordered grain boundary in copper.
In this case, $\Delta t$ is set to the maximum position $t^{*}=17$
ps on the NGP curve.

Statistical properties of strings and rings in diverse materials have
been studied in great detail \citep{Donati_1999,Douglas98,Kob_97,AlSm1,chandler2010dynamics,jung2005dynamical,10.1063/1.1644539}.
However, their origin and the formation and growth mechanisms remain
poorly understood. The MD simulations reported in this article reveal
that strings containing more than 2-4 atoms do not form by a single
coordinated/cooperative atomic displacement. Instead, such strings
represent trajectories of density fluctuations propagating through
the material by a series of discrete atomic rearrangements. Such trajectories
carry information about the relaxation dynamics of the density fluctuations.
However, the present simulations challenge the previously claimed
association of the strings with collective displacements of large
($n\gg3$) atomic groups. That association significantly overestimates
the degree of coordination among atomic movements in disordered systems.

\section{Methods\label{sec:Methods}}

The MD simulations were performed using the GPU-accelerated implementation
of the Large-scale Atomic/Molecular Massively Parallel Simulator (LAMMPS)~\citep{Plimpton95}.
Visualization and analysis of atomic configurations utilized the Open
Visualization Tool (OVITO)~\citep{Stukowski2010a}. The MD integration
time step was set to 0.1 fs. Atomic interactions were described using
the embedded atom method interatomic potential for Cu \citep{Mishin01},
the modified Tersoff potential for Si~\citep{Purja-Pun:2017aa},
the angular embedded atom potential for Al\textendash Si~\citep{Saidi_2014},
and angular-dependent interatomic potential for the Cu\textendash Ta
system~\citep{G.P.-Purja-Pun:2015aa}.

The single-crystalline Cu simulations used a fully periodic cubic
supercell whose size varied between 3.6 nm and 10.8 nm, depending
on the simulation goal. Vacancies were created by randomly removing
atoms to achieve the target vacancy concentration. To study interstitial
defects, a single Cu or Ta interstitial atom was introduced. Multiple
interstitials were not created to avoid their clustering. The Si simulations
utilized a $5.4$ nm fully periodic cubic supercell. A single vacancy
or a single interstitial atom were introduced into the system, creating
a point-defect concentration of $1.25\times10^{-4}$. The amorphous
Si model was constructed in a $10.9$ nm supercell following the protocol
described in Ref.~\citep{Moon:2021aa}.

A Cu bicrystal with a $\Sigma29\{250\}$ tilt grain boundary was generated
using the $\gamma$-surface method~\citep{Mishin:1998aa}. The bicrystal
had the approximate dimensions of $15\times15\times30$ nm$^{3}$.
Periodic boundary conditions were applied in lateral directions parallel
to the boundary plane, while free surfaces were imposed along the
normal direction. To preserve the periodicity of the grain boundary
structure, the system dimensions were slightly adjusted to ensure
an integer number of unit cells within the simulation box. A Si bicrystal
containing a $\Sigma85\{100\}$ twist grain boundary had similar boundary
conditions and was also generated using the $\gamma$-surface method~\citep{Mishin:1998aa}.
The model dimensions were $14\times14\times32.5$ nm$^{3}$. The atomic
displacements were studied in the NVT ensemble.

An Al(110)/Si(001) interphase boundary was constructed by a vapor
deposition simulation. A layer of Al$_{0.905}$(Si$_{0.095}$) solution
was deposited onto a Si(001) substrate from a gas phase at the temperature
of 648 K. The substrate dimensions were $21.5\times21.5\times7.5$
nm$^{3}$. Details of the vapor deposition can be found in Ref.~\citep{li2024atomistic}.
The layer composition and temperature were chosen so as to keep the
alloy within the single-phase region on the left of the solvus line
on the Al-Si phase diagram \citep{Al-Si-IPB} computed with the interatomic
potential. Diffusion processes were investigated in the NVT ensemble.

A Cu(110)/Ta(110) interphase boundary was created by bonding Cu and
Ta single crystals with the desired crystallographic orientations.
The system was equilibrated by semi-grand canonical Monte Carlo simulations
using the parallel Monte Carlo code ParaGrandMC \citep{ParaGrandMC}.
Careful consideration was given to minimizing the lattice mismatch
between the two bonded materials. The cross-sectional area of the
interface was $16.2\times17.8$ nm$^{2}$. Periodic boundary conditions
were applied in the in-plane directions with surface boundary conditions
normal to the interface. The system was equilibrated by isothermal
anneals at various temperatures. Interface diffusion was studied by
NVE MD simulations with various durations up to 20~ns.

\section{Results}

\subsection{Strings in structurally ordered systems}

\begin{figure}
\begin{centering}
\includegraphics[width=1\textwidth]{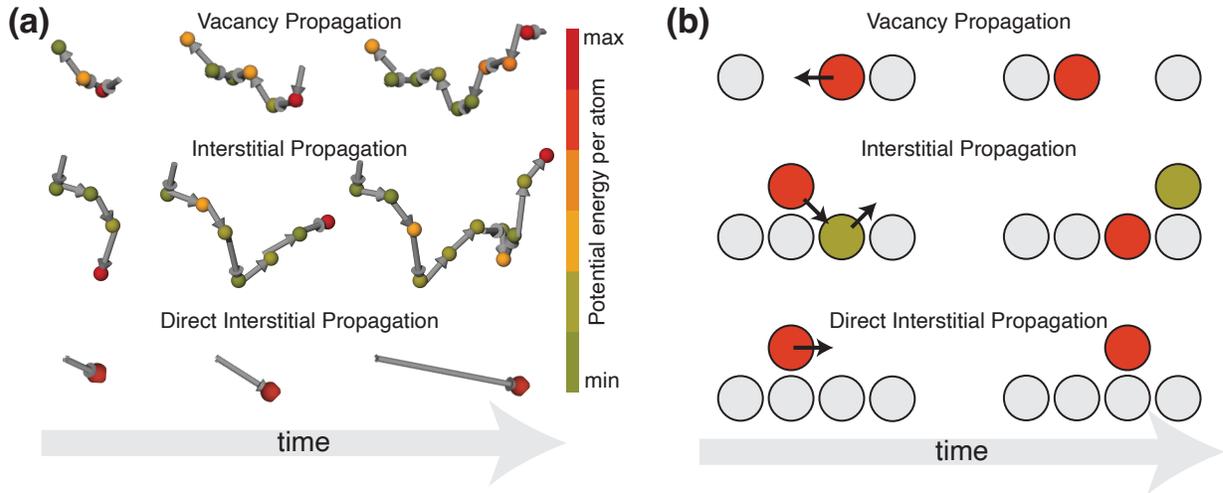}
\par\end{centering}
\caption{Strings in crystalline solids. (a) String-like trajectories of atomic
displacements in the copper lattice at 1000 K caused by vacancy diffusion
and interstitial diffusion by the indirect mechanism. For comparison,
the bottom row shows displacements of a Ta atom diffusing in Cu at
1200 K by the direct interstitial mechanism. In the latter case, the
displacements do not form a string. The snapshots are shown at three
sequential moments in time from left to right. The atoms are color-coded
by the potential energy to show that the head atom of the string has
an excess energy (red color). (b) A conceptual diagram illustrating
the backward propagation of atomic displacement by the vacancy mechanism
and forward propagation in the indirect and direct interstitial mechanisms.
Some of the atoms are colored for easier tracking. \label{fig:lattice_1}}
\end{figure}

To set the stage for the analysis of atomic dynamics in disordered
structures, we start by examining the atomic trajectories during self-diffusion
in crystalline solids mediated by vacancies and interstitial atoms.
Fig.~\ref{fig:lattice_1}(a) illustrates typical chains of atomic
displacements caused by these defects in crystalline Cu. Note that
these chains look very similar to the strings widely observed in structurally
disordered systems. The images are snapshots of the strings at sequential
points in time, showing their growth direction. Each string has a
head (the last atom joining the string in the growth direction) and
a tail (the first atom that initiated the string). If the string is
created by a vacancy, the individual atomic displacements are opposite
to the growth direction. If the string is created by an interstitial,
the atomic displacements are in the growth direction. We call these
growth mechanisms backward propagation and forward propagation, respectively.

It is important to note that the vacancy and the interstitial execute
an uncorrelated random walk by a Markov process. After each defect
jump, the system has plenty of time to thermalize (de-phase) the atomic
trajectories, erasing all memory of all previous jumps. This process
is typically described in terms of the transition-state theory \citep{Vineyard:1957vo}.
Thus, the strings shown in Fig.~\ref{fig:lattice_1}(a) represent
the atomic trajectories caused by the point defect migration. The
atoms forming the strings are almost indistinguishable from the surrounding
atoms, except for the head atom, which possess excess energy due to
its proximity to the defect. This is evident in Fig.~\ref{fig:lattice_1}(a),
where the atoms are color-coded by the potential energy. In other
words, the physically distinct part of a string is its moving head.
The rest of the string has had sufficient time to undergo nearly complete
structural relaxation and is revealed in MD simulations due to the
computer's ability to label and track individual atoms.

Note that a point defect is associated with a positive (interstitial)
or negative (vacancy) local deviation of the atomic density from the
background value. We refer to both types of the local excess density
as a ``densiton'', which can be either positive (e.g., interstitial)
or negative (e.g., vacancy). While this term is redundant in the context
of crystalline solids, it facilitates the subsequent discussion of
strings in disordered systems, in which vacancies and interstitials
are not well-defined while the respective density fluctuations play
similar roles.\footnote{The term ``densiton'' was previously mentioned in the context of sound-like
excitations in ionic liquids \citep{Apostol:2008aa}. The densitons
introduced here have a fundamentally different meaning.} In this terminology, we can say that the strings shown in Fig.~\ref{fig:lattice_1}(a)
are trajectories of the positive and negative densitons.

A moving positive densiton causes a forward propagation of atomic
displacements, while a moving negative densiton causes a backward
propagation. Each displacement of a negative densiton is accomplished
by an atomic jump in the opposite direction, filling the low-density
site and creating a new low-density site (densiton) behind. This is
the mechanism of the backward propagation. A displacement of a positive
densiton can be a more complex process. At relatively low temperatures,
interstitial atoms in Cu form a split dumbbell configuration parallel
to $\left\langle 100\right\rangle $ directions. A dumbbell jumps
by a concerted displacement of three atoms moving the dumbbell center
into a nearest-neighboring lattice site with a simultaneous $90^{\circ}$
rotation. Thus, the forward propagation caused by a positive densiton
occurs by a chain of atomic rearrangements involving a collective
displacement of three atoms. At high temperatures, however, thermal
fluctuations nearly destroy the close coordination of the three atoms,
and the process is better described by a chain of individual atomic
jumps.

It is important to recognize that the forward propagation of a string
is only possible due to the indirect interstitial (also called interstitialcy)
mechanism, in which each displacement of a positive densiton moves
a new group of atoms. As a result, each atom in a long string only
undergoes a nearest-neighbor displacement. The direct interstitial
mechanism, in which a single interstitial atom hops between interstitial
positions, does not produce a string. This is illustrated in Fig.~\ref{fig:lattice_1}(a)
for a Ta interstitial atom in Cu. Instead of a string, the MD simulation
produces a single long arrow pointing to the time-dependent position
of the Ta atom. The three atomic mechanisms discussed here are summarized
in the conceptual diagram in Fig.~\ref{fig:lattice_1}(b).

Finally, note that the strings in crystalline systems can grow to
a very long but finite size. Several mechanisms can stop the growth.
A winding trajectory can accidentally cross itself, breaking the string
into two parts. Furthermore, if two different trajectories accidentally
cross, each of them will be chopped into two parts. In a relatively
small system, the maximum string length is on the order of the system
size. If the crystal contains both vacancies and interstitials, they
can mutually annihilate, terminating two strings.

To summarize the above analysis, the standard algorithm based on the
$\Delta t=t^{*}$ time interval identifies relatively long strings
comprising $n\gg3$ atoms. Such strings can rarely form by a single
collective event. They represent a chain of dynamically uncorrelated
atomic rearrangements typically involving from 1 to 3 atoms. A string
grows by the motion of a densiton located at the string's head. The
string represents the densiton's trajectory, which is physically almost
indistinguishable from the background except for its head. Depending
on the densiton's sign, the growth occurs by either forward or backward
propagation of discrete atomic displacements. The string growth continues
as long as the densiton exists. The observation of a moving positive
densiton indicates that the atomic displacements propagate by an indirect
mechanism.

\subsection{Strings in structurally disordered systems}

\begin{figure}
\begin{centering}
\includegraphics[width=1\textwidth]{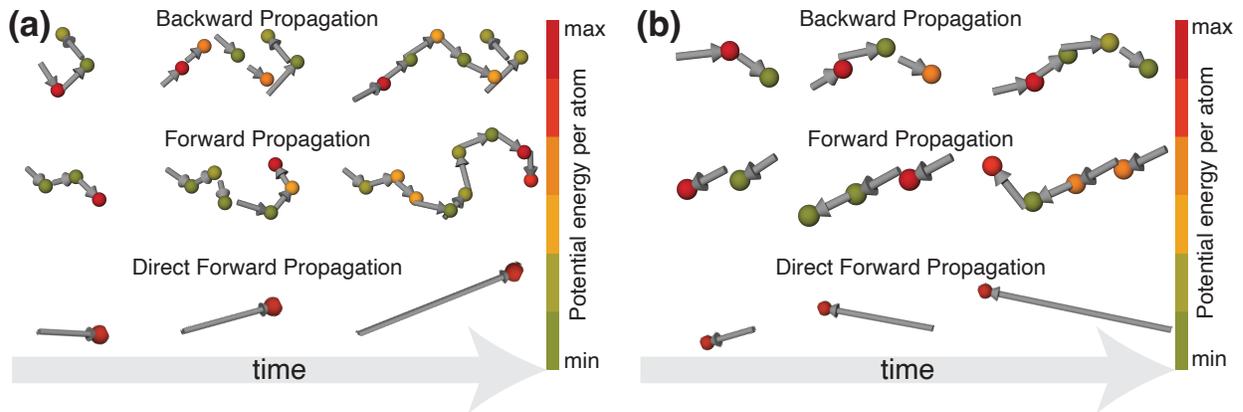}
\par\end{centering}
\caption{Examples of strings in disordered structures. (a) Backward and forward
propagating strings in (a) $\Sigma29$ $(250)$ tilt grain boundary
in Cu at 1000 K and (b) undercooled liquid Cu at 1000 K. In both cases,
the lower row presents an example of direct forward propagation of
a single atom that does not create a string. The snapshots are shown
at three sequential moments in time from left to right. The atoms
are color-coded by the potential energy. \label{fig:Disordered_1}}
\end{figure}

We next consider strings in structurally disordered systems. We have
studied the formation and growth of strings and rings in undercooled
liquids and thermally disordered interfaces such as grain boundaries
and interphase boundaries. The materials that we studied included
elemental Cu and Si, as well as Al-Si interfaces obtained by simulated
vapor deposition and Cu-Ta interfaces obtained by the direct bonding
method. Most of the results are presented in the Supplementary Information
file. Representative examples are shown in Fig.~\ref{fig:Disordered_1}
for strings in a Cu grain boundary and in undercooled Cu liquid. Note
the striking similarity with the strings produced by vacancy and interstitial
migration in crystalline Cu (cf.~Fig.~\ref{fig:lattice_1}(a)).
In both ordered and disordered systems, some strings grow by forward
propagation of atomic jumps while others grow by backward propagation.
In all cases studied here, the atoms forming a string could not be
physically distinguished from the surrounding atoms in the MD simulations.
(In fully disordered structures, even the moving head of a string
could not be reliably identified by our image analysis if the arrows
showing the atomic displacements were erased.)

This remarkable similarity strongly suggests that the atomic dynamics
in disordered systems are governed by fundamentally similar atomic-level
mechanisms to those in crystals. The concept of a vacancy in disordered
structures is not well-defined due to the absence of an underlying
lattice. However, the observation of back-propagating strings suggests
the existence of similar local-low-density structures that can migrate
by a vacancy-like mechanism. Namely, one of the nearby atoms jumps
into the low-density ``cage'' and fills it, leaving a similar low-density
spot behind. Occasionally, a group of $n_{c}=2$ to $4$ atoms can
participate in the jump. Repeated atomic rearrangements of this type
create a chain of atomic displacements in the direction opposite to
the low-density propagation (Fig.~\ref{fig:lattice_1}(a)). Recognizing
that such vacancy-like low-density structures are not ``real'' vacancies,
we unify both crystalline and disordered cases under the term of a
negative densiton.

Similarly, in the absence of a crystalline lattice, interstitials
in disordered structures are not well-defined. However, the existence
of forward-propagating strings is strong evidence that disordered
structures can host similar local-high-density excitations, which
we call positive densitons. They can migrate through the system by
causing a chain of relatively short ($\approx r_{0}$) atomic displacements
in the forward direction, creating a forward-propagating string. This
mechanism is similar to the indirect interstitial migration mechanism
in crystalline solids. Other authors also suggested the existence
of a high concentration of interstitial-like ``defects'' in liquid
phases with properties similar to those of real interstitials existing
in the respective solid phase near the melting point \citep{Nordlund2005,Ashkenazy2010}.

\begin{figure}
\begin{centering}
\includegraphics[width=1\textwidth]{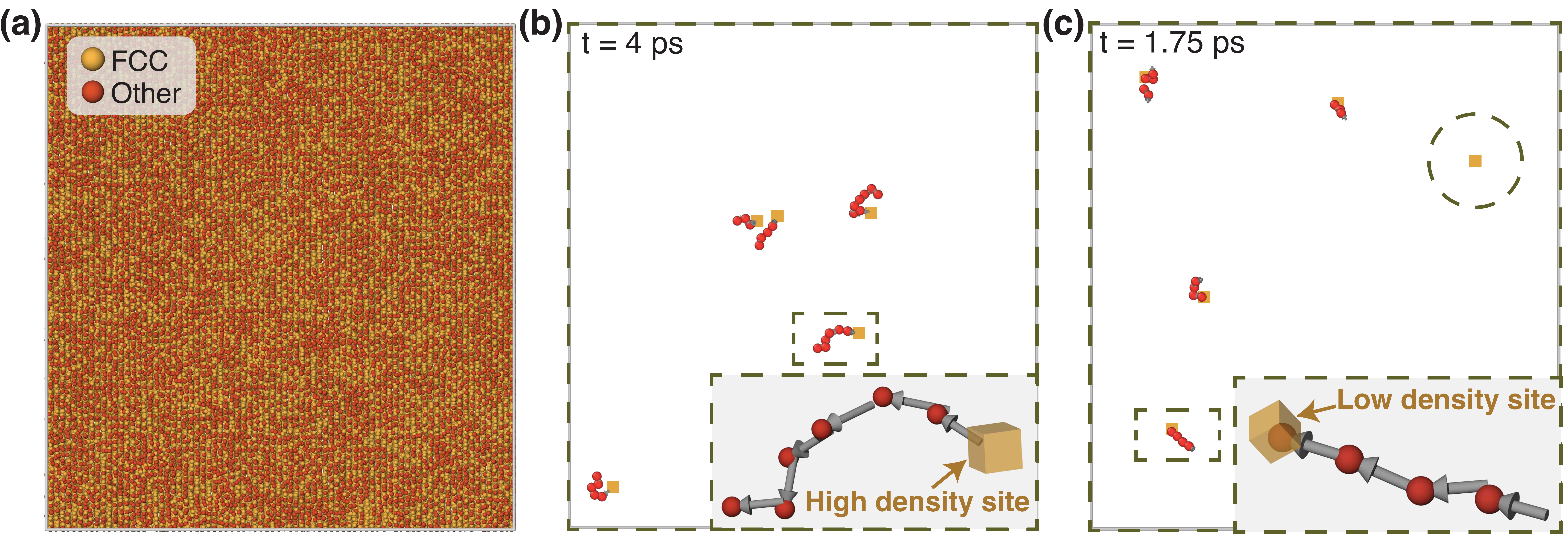}
\par\end{centering}
\caption{String nucleation at implanted densitons. (a) Top view of the $\Sigma29$
$(250)$ tilt grain boundary in Cu at 1000 K, showing its disordered
structure. Atoms in face-centered cubic (FCC) environments are colored
in yellow, while all other atoms are colored red. (b) Five atoms were
randomly inserted into the boundary, creating high-density sites (positive
densitons). After a 4 ps NVE MD run, all five densitons gave rise
to forward-propagating strings. One of them is shown in the inset.
(c) Five random atoms were removed from the boundary, creating low-density
sites (negative densitons). After a 1.75 ps NVE MD run, four of them
gave rise to backward-propagating strings. One of them is shown in
the inset. One low-density site (encircled in the top right corner)
delocalized without creating a string. \label{fig:Disordered_2}}
\end{figure}

The link between strings and densitons is further demonstrated in
Fig.~\ref{fig:Disordered_2}. In one test, a few atoms were randomly
inserted into a disordered grain boundary, creating a set of high-density
spots. In another test, a few atoms were randomly removed from the
same grain boundary, creating a set of low-density spots. In other
words, we have artificially planted a set of positive or negative
densitons into the boundary. In the subsequent MD simulations, most
of the positive densitons drifted away from the initial location and
created a forward-propagating string. Likewise, most of the negative
densitons gave rise to a backward-propagating string. These tests
demonstrate that densitons can persist for an extended period of time
and migrate by creating strings. In some cases, a negative densiton
would not produce a string and would, instead, dissolve in the surrounding
structure (Fig.~\ref{fig:Disordered_2}(c)). Such cases provide examples
of short-lived densitons whose lifetime is shorter than the inverse
frequency of atomic jumps during the string growth.

Similar scenarios were earlier observed by Delaye and Limoge \citep{Limoge-1993},
who examined static structural relaxations in a Lennard-Jones glass
after removing an atom. They found that the displacements of neighboring
atoms could be large enough to fill the void and restore the average
atomic density. In other cases, the displacements could be small enough
to preserve the local excess free volume. In yet another scenario,
the relaxation had the form of a ``cooperative evolution'' involving
many atoms. The latter process apparently represented a string nucleation,
although its nature was not studied in detail.

\subsection{Exploring the timescales of string evolution}

\begin{figure}
\includegraphics[width=1\textwidth]{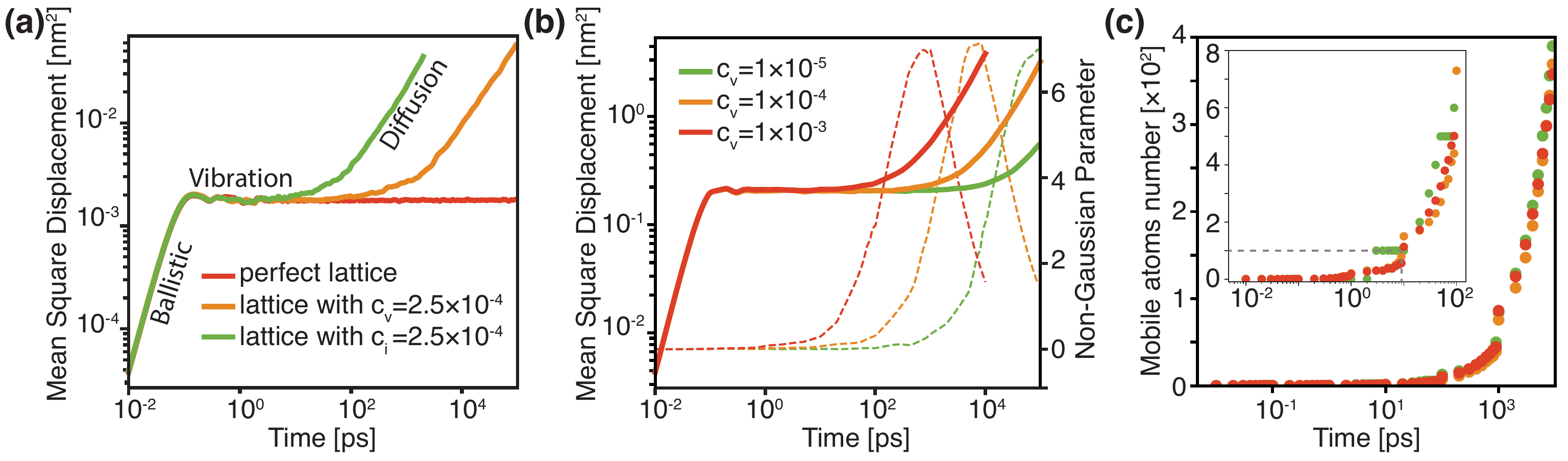}

\caption{(a) A log-log plot of the mean square displacement (MSD) versus time
for copper at 1000 K, comparing a perfect lattice with a lattice containing
one vacancy ($v$) or one interstitial atom ($i$). The respective
defect concentrations (per lattice site) are $c_{v}=c_{i}=2.5\times10^{-4}$.
The plot exhibits the ballistic, vibrational, and diffusion regimes
of atomic motion. (b) Similar plot for Cu lattice with three different
vacancy concentrations, with superimposed NGP plots for the respective
concentrations. (c) Number of mobile atoms normalized by the number
of vacancies versus the time interval $\Delta t$ for the three vacancy
concentrations shown in (b). The inset is a zoom into the time interval
with a relatively small number of mobile atoms per vacancy.\label{fig:lattice-MD}}
\end{figure}

The string growth involves several different time scales. We will
discuss them starting with a crystalline system. Fig.~\ref{fig:lattice-MD}
displays several log-log plots of the MSD versus time $\Delta t$
over several orders of magnitude. For the defect-free Cu lattice,
the plot is linear with a slope of two in the short time limit. This
linear behavior represents the ballistic motion of the atoms. At $\Delta t\approx10^{-1}$
ps, the plot reaches a local maximum followed by a few oscillations,
signaling a dynamic transition from the ballistics to atomic vibrations.
The time interval of this transition is approximately $10^{-1}\lesssim\Delta t\lesssim1$
ps, which is comparable to the inverse of the Debye frequency of Cu
($\nu_{D}=7.14\times10^{12}$ Hz at 0 K estimated from $\nu_{D}=k_{B}T_{D}/h$
using the Debye temperature of $T_{D}=343$ K \citep{Kittel}). This
transition is followed by a horizontal segment representing the MSD
of atomic vibrations at 1000 K.

For systems with a single point defect, the MSD curve shows an upward
deviation from the perfect-lattice behavior when $\Delta t$ reaches
the timescale of string nucleation (i.e., activation of point-defect
jumps) (Fig.~\ref{fig:lattice-MD}(a)). This time is shorter for
an interstitial than for a vacancy due to the lower activation barrier.
The horizontal segment on the curve evolves into a shoulder. Since
point-defect jumps are dynamically memoryless, the jump activation
time is the same as the average time increment of string growth, which
we call the jump waiting time $\tau_{w}$. To quantify this time for
vacancy-induced strings, we computed MSD curves for three vacancy
concentrations. As expected, the vacancy-induced deviation of the
MSD curve from the lattice behavior starts earlier as the vacancy
concentration increases (Fig.~\ref{fig:lattice-MD}(b)). Note, however,
this shift of the curves does not mean that the string activation
time $\tau_{w}$ becomes shorter. Since the vacancies do not interact
with each other, $\tau_{w}$ does not change. This example demonstrates
that the location of the shoulder on the MSD curves does not characterize
the string activation time.

To demonstrate this point further, we have computed the average number
$N_{vj}$ of vacancy jumps in the system as a function of time for
the three vacancy concentrations. While $N_{vj}$ increases with the
number $N_{v}$ of vacancies present in the system, the number of
jumps per vacancy, $N_{vj}/N_{v}$, does not change. In other words,
the $N_{vj}/N_{v}$ versus time plots for different $N_{v}$ collapse
into a single master curve representing the true dynamics of a single
vacancy (Fig.~\ref{fig:lattice-MD}(c)). Using this master curve,
we can estimate $\tau_{w}$ as the time for which $N_{vj}/N_{v}=1$,
which is a few picoseconds.

Returning to Fig.~\ref{fig:lattice-MD}(b), note that the NGP maximum
occurs on a significantly longer timescale than $\tau_{w}$. On this
timescale, a vacancy can make over 10 jumps. Moreover, the NGP peak
position shifts with the vacancy concentration. Given the previously
discussed similarity between the ordered and disordered system, the
following conclusions can be made: (1) the time $t^{*}$ commonly
used for string identification is too long ($\gg\tau_{w}$) and overlooks
the key mechanisms of the string growth, and (2) $t^{*}$ reflects
not only the growth of individual strings but also the degree of disorder
in the system. In our terminology, $t^{*}$ depends not only on the
densiton dynamics but also on their concentration. Disentangling the
two factors is not an easy task.

\begin{figure}
\includegraphics[width=1\textwidth]{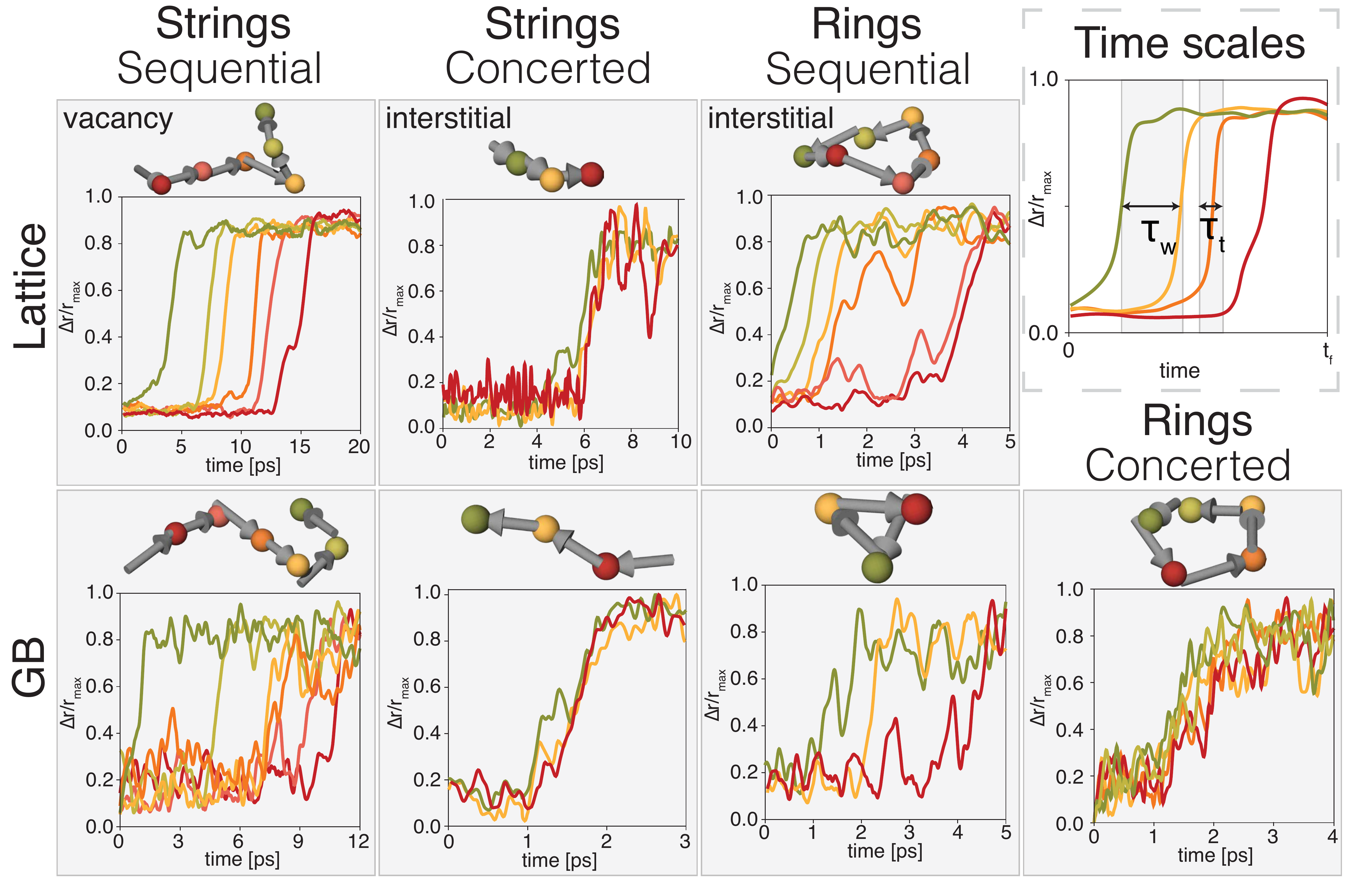}

\caption{Examples of strings and rings formed by sequential and concerted atomic
rearrangements in Cu lattice and $\Sigma29$ $(250)$ tilt grain boundary
(GB). The temperature is 1000\,K for all cases except for the concerted
atomic rearrangement caused by an interstitial dumbbell in the Cu
lattice at 500 K. The plots show displacements of individual atoms
as a function of time interval $\Delta t$. The displacements are
normalized by their maximum value during the period $\Delta t$. The
curves have been smoothed using a running average to eliminate the
noise due to vibrational motion. The schematic in the top right corner
explains the definitions of the the transition/rearrangement time
$\tau_{t}$ and the waiting time $\tau_{w}$.\label{fig:Strings-rings}}
\end{figure}

Another relevant timescale is one on which the atomic rearrangements
occur at the head of the string. The rearrangement/transition time,
which we denote $\tau_{t}$, is significantly shorter than the waiting
time $\tau_{w}$. For lattice strings, $\tau_{t}$ is roughly on the
order of $1/\nu_{D}$. For concerted atomic rearrangements involving
several atoms, $\tau_{t}$ is somewhat longer but still below $\tau_{w}$.
The diagram in the top right corner of Fig.~\ref{fig:Strings-rings}
explains the difference between $\tau_{w}$ and $\tau_{t}$ using
a plot of atomic displacements versus time for a set of atoms involved
in a string. This graphical format \citep{10.1063/1.1644539,annamareddy_superionic,Erhart2020}
allows one to distinguish between sequential atomic jumps and truly
collective events. In the latter case, several curves bunch together
into a single transition. Fig.~\ref{fig:Strings-rings} summarizes
such plots for both sequential and concerted atomic rearrangements
in ordered and disordered systems, highlighting the differences between
$\tau_{w}$ and $\tau_{t}$. In addition to open strings, Fig.~\ref{fig:Strings-rings}
shows rings, which are observed less frequently. Like open strings,
rings can also evolve by either a sequence of single-atom jumps or
as a single concerted displacement of an atomic group. The bottom
right corner in Fig.~\ref{fig:Strings-rings} shows a rare case of
a five-atom ring in a grain boundary, which forms by a single concerted
atomic displacement.

\begin{figure}
\begin{centering}
\includegraphics[width=0.6\textwidth]{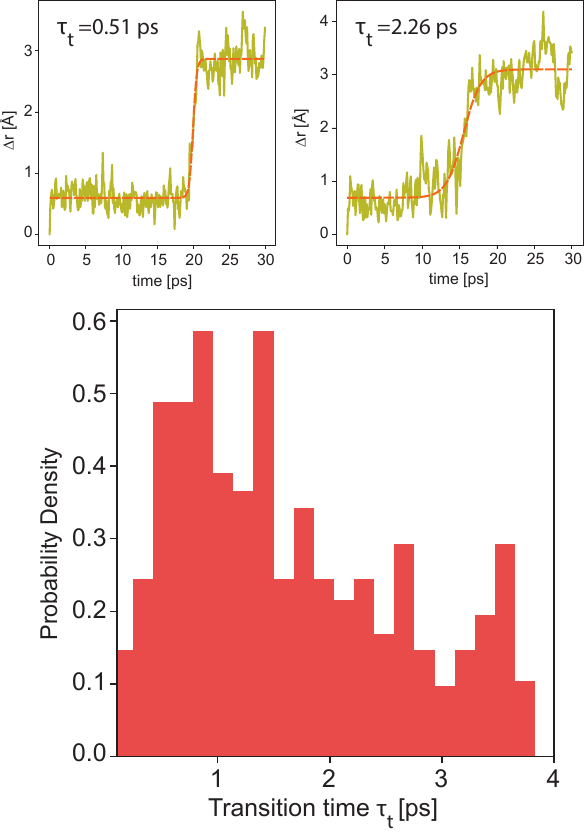}
\par\end{centering}
\caption{Evaluation of the collective transition timescale $\tau_{t}$ from
MD simulations of the disordered $\Sigma29$~\{250\} grain boundary
in Cu at 800 K. (a,b) Representative displacement-time plots for single-atom
transitions and their fits with a $\tanh(t/\tau_{t})$ function (dashed
lines) to extract the $\tau_{t}$ values. (a) Example of small $\tau_{t}$.
(b) Example of large $\tau_{t}$. (c) Probability distribution of
$\tau_{t}$ values.\label{fig:Evaluation-of-tau}}
\end{figure}

To quantify the timescales $\tau_{w}$ and $\tau_{t}$ more precisely
and estimate the number $n_{c}$ of atoms involved in collective transitions,
consider the disordered $\Sigma29$~\{250\} grain boundary in Cu
as a representative case. This boundary already appeared in Figs.~\ref{fig:1},
\ref{fig:Disordered_1} and \ref{fig:Disordered_2}. Additional MD
runs were performed for this boundary in the NVE ensemble at the temperature
of 800 K. All string-like atomic rearrangements were detected by the
method explained in section \ref{sec:Methods} and stored in the computer
memory. Collective transitions manifested themselves in sudden jumps
in displacement-time plots, as was shown in Fig.~\ref{fig:Strings-rings}.
Statistics were collected for the durations $\tau_{t}$ of such jumps.
To estimate $\tau_{t}$, the jump event was fitted by a $\tanh(t/\tau_{t})$
function, as illustrated in Fig.~\ref{fig:Evaluation-of-tau}(a,b).
The statistical distribution of $\tau_{t}$ peaks between 1 and 2
ps. In the example shown in Fig.~\ref{fig:Evaluation-of-tau}(c),
the peak is at $\tau_{t}\approx1.6$ ps. Based on these statistics,
we take $\tau_{t}=2$ ps as a reasonable estimate of $\tau_{t}$ for
this grain boundary.

\begin{figure}
\begin{centering}
\includegraphics[width=0.53\textwidth]{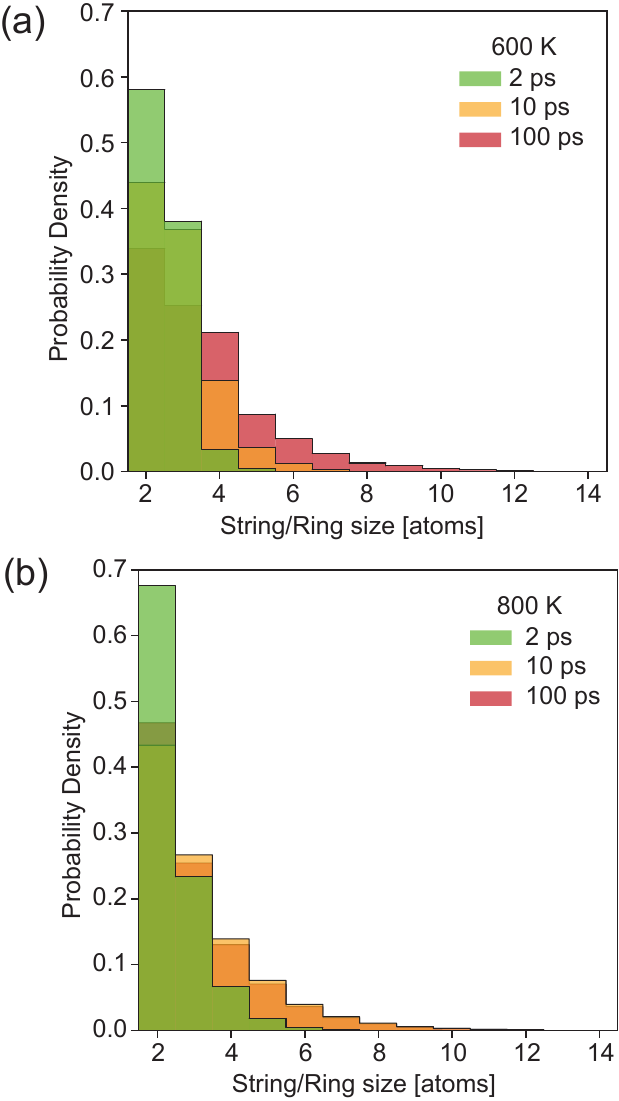}
\par\end{centering}
\caption{Evaluation of the number of atoms $n_{c}$ participating in collective
transitions. The disordered $\Sigma29$~\{250\} grain boundary in
Cu at (a) 600 K and (b) 800 K is used as an example. The statistical
distributions of the string size are shown for three values of the
time window $\Delta t$ indicated in the key. The window $\Delta t=\tau_{t}$
detects predominantly collective atomic rearrangements.\label{fig:histograms}}
\end{figure}

To estimate the number of atoms $n_{c}$ participating in collective
rearrangements, we scanned the history of a 100 ps long MD run with
a moving time window of width $\Delta t$=$\tau_{t}$, counting the
number of atoms undergoing string-like rearrangements within this
window. By the meaning of $\tau_{t}$, such rearrangements are collective
in nature. Fig.~\ref{fig:histograms} shows the obtained statistical
distributions of $n_{c}$ values for the temperatures of 600 K and
800 K. At 600 K, 56\% of the collective events involve $n_{c}=2$
atoms, another 38\% involve $n_{c}=3$ atoms, and only 3.5\% of the
collective events involve $n_{c}=4$ atoms. At 800 K, the respective
percentages are 68\%, 23\%, and 6.7\%. The shape of the distribution
changes with temperature, but the conclusion remains that the overwhelming
majority of the collective rearrangements involve between 2 and 3
atoms. Fig.~\ref{fig:histograms} additionally displays the histograms
for much wider time windows $\Delta t$. Note that the string size
increases with $\Delta t$ because the wider time windows ($\Delta t\gg\tau_{t}$)
capture longer strings formed by sequences of several collective events.

\begin{figure}
\begin{centering}
\includegraphics[width=0.55\textwidth]{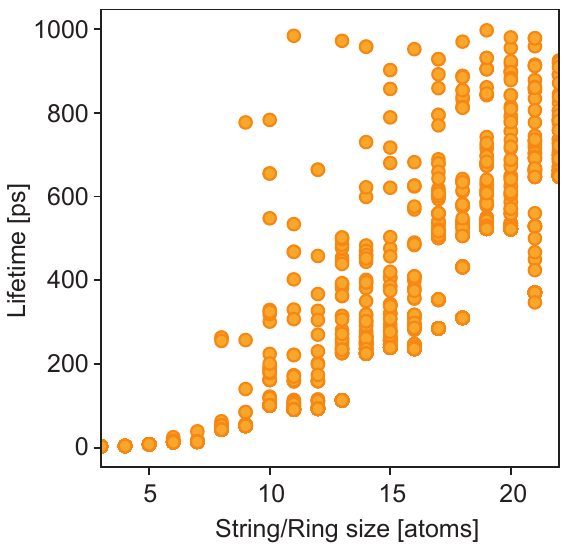}
\par\end{centering}
\caption{The string lifetime as a function of string length $n$ in the disordered
$\Sigma29$~\{250\} grain boundary in Cu at 600 K. \label{fig:lifetimes}}
\end{figure}

As was illustrated in Fig.~\ref{fig:Strings-rings}, the waiting
time $\tau_{w}$ between the collective events is significantly longer
than $\tau_{t}$. This is further confirmed by the plot of the string
lifetime versus the string size $n$ in Fig.~\ref{fig:lifetimes}.
The string lifetime was extracted from the simulation history by detecting
the time when the string first appeared and the time after which it
no longer grew. The large scatter of the points in the long-time limit
reflects the broader distribution of the lifetimes of longer strings.
The plot shows that it takes hundreds of picoseconds to grow a string
composed of 10 or more atoms. A 10-atom string could be formed by
3 to 5 back-to-back atomic rearrangements, which would only take 6
to 10 ps. The fact that it actually takes $\sim100$ ps to grow such
a string confirms the intermittent nature of the string growth with
long ($\tau_{w}\gg\tau_{t}$) time intervals separating the collective
events.

It should be noted that the above estimates of $n_{c}$ and the timescales
$\tau_{w}$ and $\tau_{t}$ were obtained for a particular (although
rather broad) set of disordered systems at relatively high temperatures.
We did not consider other disordered systems, such as metallic glasses,
at low temperatures. In such systems, the numerical values of the
above parameters could be different, although the described mechanism
of string growth is likely to remain qualitatively similar. The timescale
of collective atomic rearrangements and the dephasing time after each
collective event should still be in the picosecond range (roughly,
several vibration periods). More atoms could be involved in the collective
events than at high temperatures, but the difference is unlikely to
be dramatic. The densiton lifetime can be longer, resulting in longer
strings. But the strings should still grow intermittently with relatively
long ($\gg1$ ps) periods between fast ($\sim1$ ps) collective rearrangements.
These predictions require validation by direct MD simulations in the
future.

\section{Discussion and conclusions}

Previous MD simulations of disordered systems were focused on collective
statistical metrics characterizing the atomic dynamics on diverse
length and time scales. The statistical methods employed included
the analysis of MSD-time functions \citep{Donati_1999,Berthier2011,Douglas98,Kob_97,AlSm1,10.1063/1.1644539},
the van Hove spatial correlation function \citep{10.1063/1.1644539,Berthier2011,Derlet:2021aa,AlSm1},
the self-intermediate scattering functions \citep{AlSm1,10.1063/1.1644539},
the size distributions of dynamics clusters \citep{Donati_1999,Douglas98,Kob_97,AlSm1,chandler2010dynamics,jung2005dynamical,10.1063/1.1644539,Karmakar:2009aa,Cui:2001aa},
and other statistical measured demonstrating dynamic heterogeneity.
We are aware of only two prior investigations of atomic displacements
in individual strings \citep{10.1063/1.1644539,Erhart2020}. In \citep{10.1063/1.1644539},
the displacements were found to propagate sequentially with occasional
collective jumps of small atomic groups called ``micro-strings''.
However, the string was treated as a physically distinct dynamical
object with a ``backbone'' hosting the micro-strings. Furthermore,
the previous studies identified the strings using a timescale $\Delta t\approx t^{*}$
on which the strings had grown relatively long and were unlikely to
have formed by a single concerted rearrangement of the participating
atoms. Because the entire string emerged in the simulation during
the time increment $\Delta t$, this gave the impression that the
atoms had moved collectively.

In the present study, we have focused on a detailed analysis of individual
atomic trajectories with small ($\ll\tau_{t}$) time increments. We
find that strings containing more than a few atoms do not form by
a single collective rearrangement. Instead, they grow by a sequence
of dynamically uncorrelated increments. Each increment occurs by a
local atomic rearrangement at the head of the string. The rearrangement
involves either a single-atom jump or a concerted displacement of
2-3 atoms. In either case, the local rearrangement happens very fast
(on the $\tau_{t}$ timescale) while the waiting time $\tau_{w}$
between sequential rearrangements is significantly longer. Importantly,
the part of the string left behind the moving head is structurally
relaxed and physically almost indistinguishable from the surrounding
atoms. It can still be revealed by tracking the atoms in MD simulations.
The moving head of the string is physically the most distinct part
due to its excess energy and (positive or negative) local free volume.

The simulations have revealed a remarkable similarity between the
strings and rings in disordered structures and the atomic trajectories
created by vacancy and interstitial migration in crystalline solids.
To reflect this similarity, we describe the string growth in both
ordered and disordered structures as migration of positive (interstitial-like)
or negative (vacancy-like) densitons. The densiton sign can be determined
in MD simulations from the direction of the atomic displacements relative
to the direction of the densiton migration (i.e., the direction of
string growth). We have shown that positive densitons migrate by forward
propagation while negative densitons migrate by backward propagation.
The strings created by moving densitons in disordered structures are
analogs to the MD trajectories of point defects in crystals.

Densitons have a distribution of lifetimes. Short-lived densitons
delocalize before they can migrate. Densitons with a longer life (persistent
densitons) migrate and can create relatively long strings before they
delocalize or encounter another densiton of the opposite sign. Interpretation
of the average string length $\overline{l}$ is not straightforward
because it involves several factors. On one hand, $\overline{l}$
depends on the densitons' lifetime, which is a physical factor. On
the other hand, MD simulations capture purely computational string-truncation
processes such as their mutual intersections and self-intersections.
Nevertheless, the observation of strings is useful as a means of locating
the densitons and establishing their sign.

The present simulations call for a significant reinterpretation of
the strings and rings in disordered structures as trajectories of
moving densitons and not dynamic objects formed by a single collective
atomic rearrangement. Instead of focusing on the strings, the atomic
dynamics can be better described by a random walk of positive and
negative densitons populating the system and exhibiting a spectrum
of mobilities and lifetimes. Fluctuations in the densiton concentration
cause the dynamic heterogeneity. The densiton migration can be responsible
for the atomic self-diffusion and possibly other transport properties
of disordered systems.

\bigskip{}

\noindent\textbf{Acknowledgments}

We are grateful to Dr.~Ian Chesser for carefully reading the manuscripts
and providing helpful suggestions. This research was supported by
the U.S. Department of Energy, Office of Basic Energy Sciences, Division
of Materials Sciences and Engineering, under Award \# DE-SC0023102.

\bigskip{}


\newpage{}

\newpage\clearpage{}

\global\long\def\thefigure{S\arabic{figure}}%
 \setcounter{figure}{0}
\begin{center}
{\LARGE\textbf{The origin of strings and rings in the atomic dynamics
of disordered systems}}{\LARGE\par}
\par\end{center}

\bigskip{}

\begin{center}
{\large Omar Hussein, Yang Li, and Y. Mishin}{\large\par}
\par\end{center}

\bigskip{}

\begin{center}
{\large Department of Physics and Astronomy, MSN 3F3, George Mason
University, Fairfax, Virginia 22030, USA}{\large\par}
\par\end{center}

\bigskip{}

\begin{center}
{\Large\textsf{\textbf{SUPPLEMENTARY INFORMATION}}}{\Large\par}
\par\end{center}

\bigskip{}

This file contains additional figures showing the investigations of
string dynamics in systems other than those discussed in the main
text.

\newpage{}

\begin{figure}[h]
\includegraphics[width=1\textwidth]{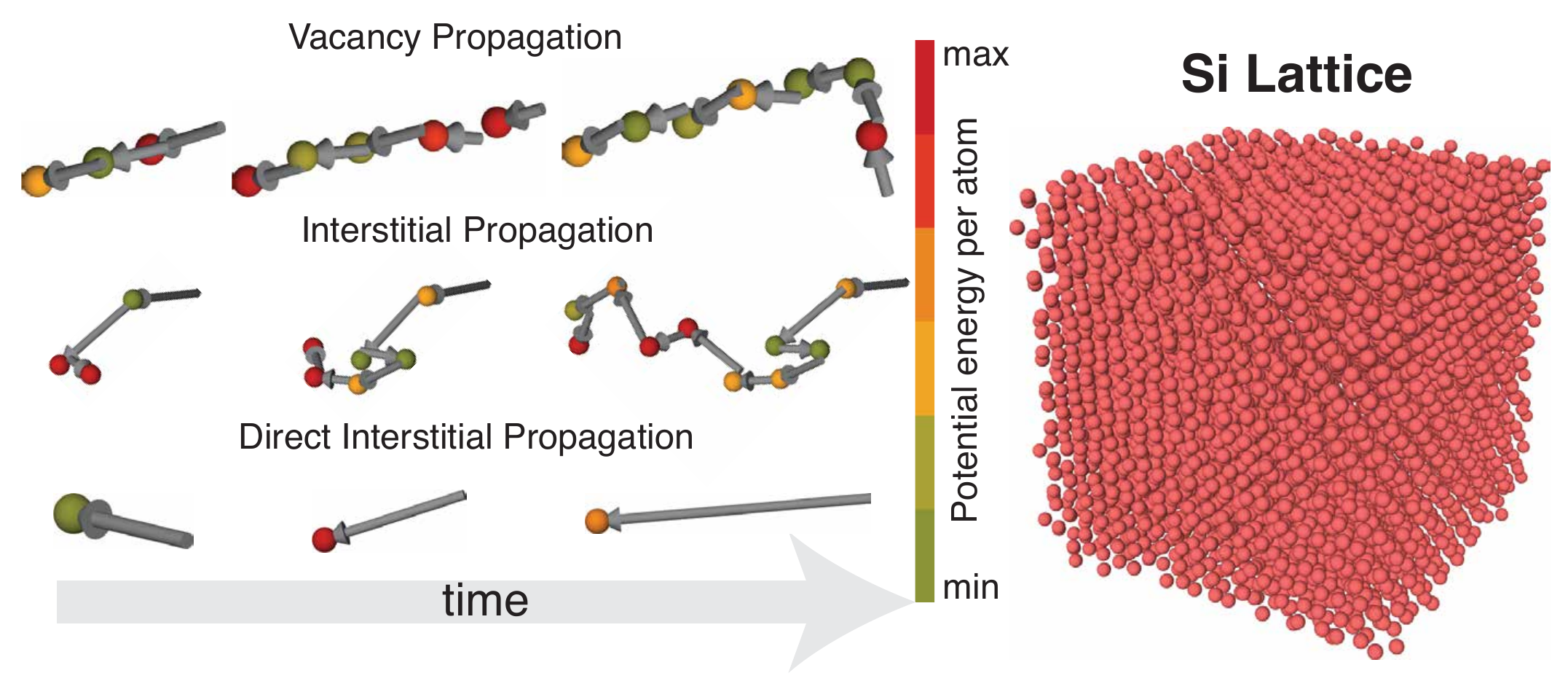}

\caption{String-like trajectories of atomic displacements in diamond-cubic
silicon lattice at 1600 K. The model has the dimensions of $5.4\times5.4\times5.4$
nm$^{3}$. Periodic boundary conditions are applied along all spatial
directions.}
\end{figure}

\bigskip{}

\bigskip{}

\begin{figure}[H]
\includegraphics[width=1\textwidth]{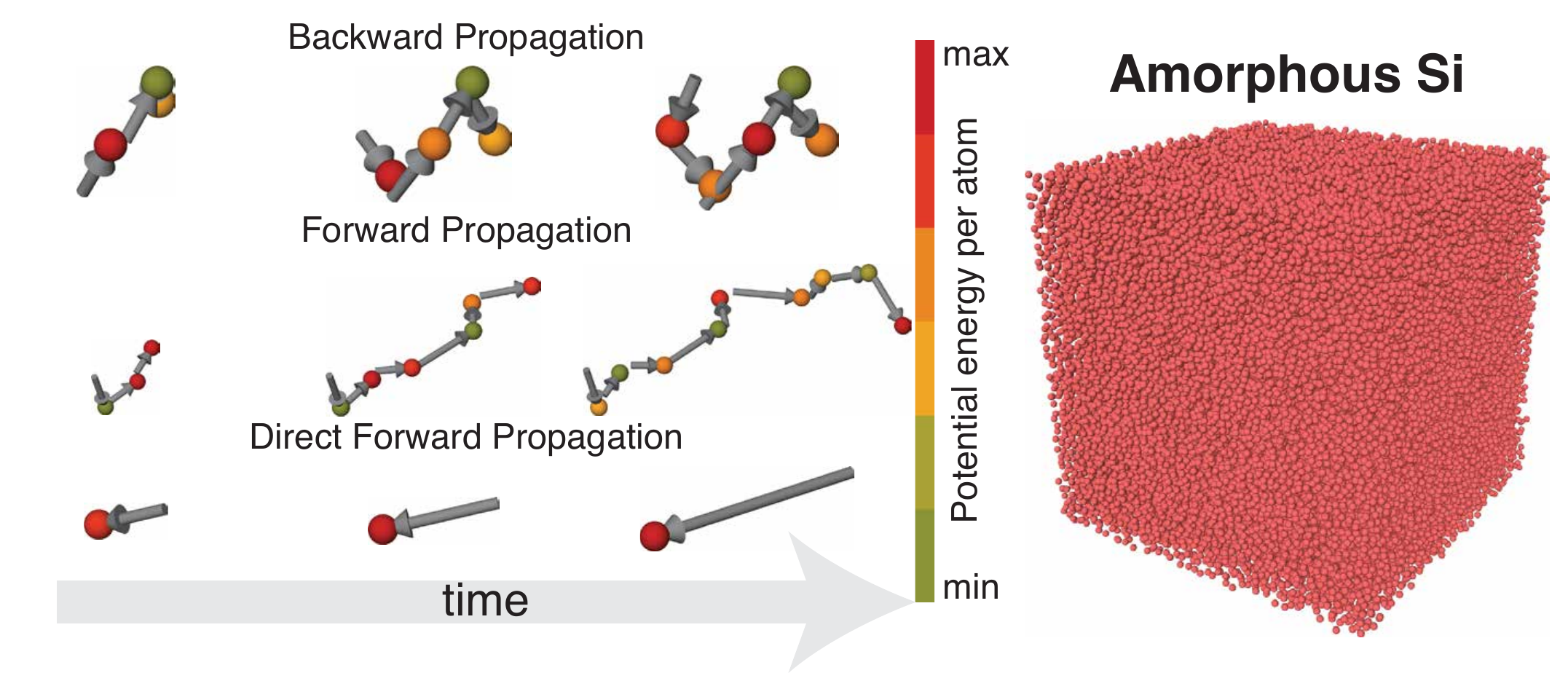}

\caption{String-like trajectories of atomic displacements in amorphous silicon
at 1000 K. The model has size of $10.9\times10.9\times10.9$ nm$^{3}$.
Periodic boundary conditions are applied along all spatial directions.
The amorphous state was prepared with the method proposed in \citep{Moon:2021aa}.}
\end{figure}

\begin{figure}[H]
\includegraphics[width=1\textwidth]{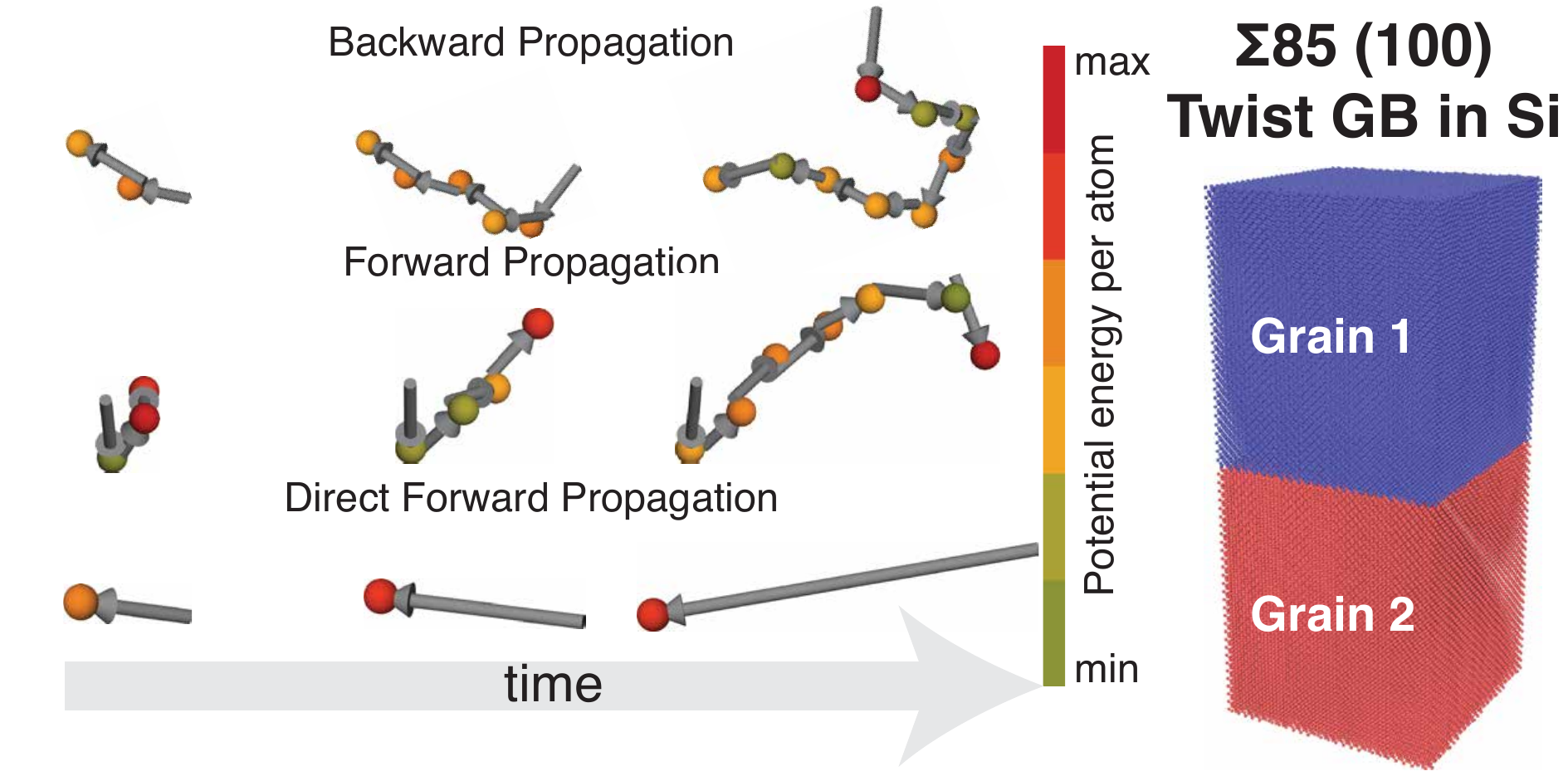}

\caption{String-like trajectories of atomic displacements at the $\Sigma$85
(100) twist grain boundary in silicon at 1600 K. The model has the
dimensions of $14\times14\times32.5$ nm$^{3}$. Periodic boundary
conditions are applied along the directions parallel to the grain
boundary plane.}
\end{figure}

\bigskip{}

\bigskip{}

\begin{figure}[H]
\includegraphics[width=1\textwidth]{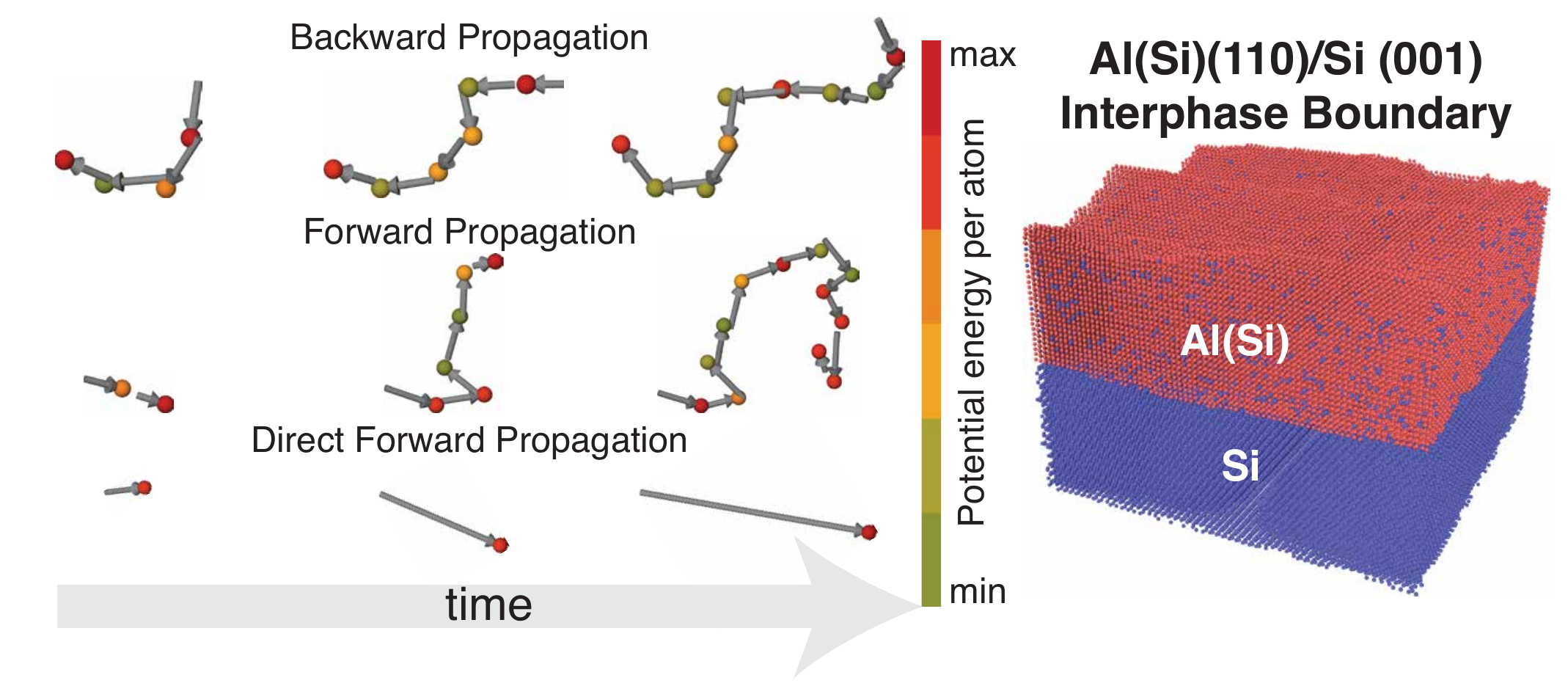}

\caption{String-like trajectories of atomic displacements at the Al(Si)(110)/Si(001)
interphase boundary in the Al/Si system at 648 K. The model has the
dimensions of $21.5\times21.5\times16$ nm$^{3}$. Periodic boundary
conditions are applied along the directions parallel to the interface.
The structure was constructed by depositing Al and Si on a Si(001)
substrate at 648 K.}
\end{figure}

\begin{figure}[H]
\includegraphics[width=1\textwidth]{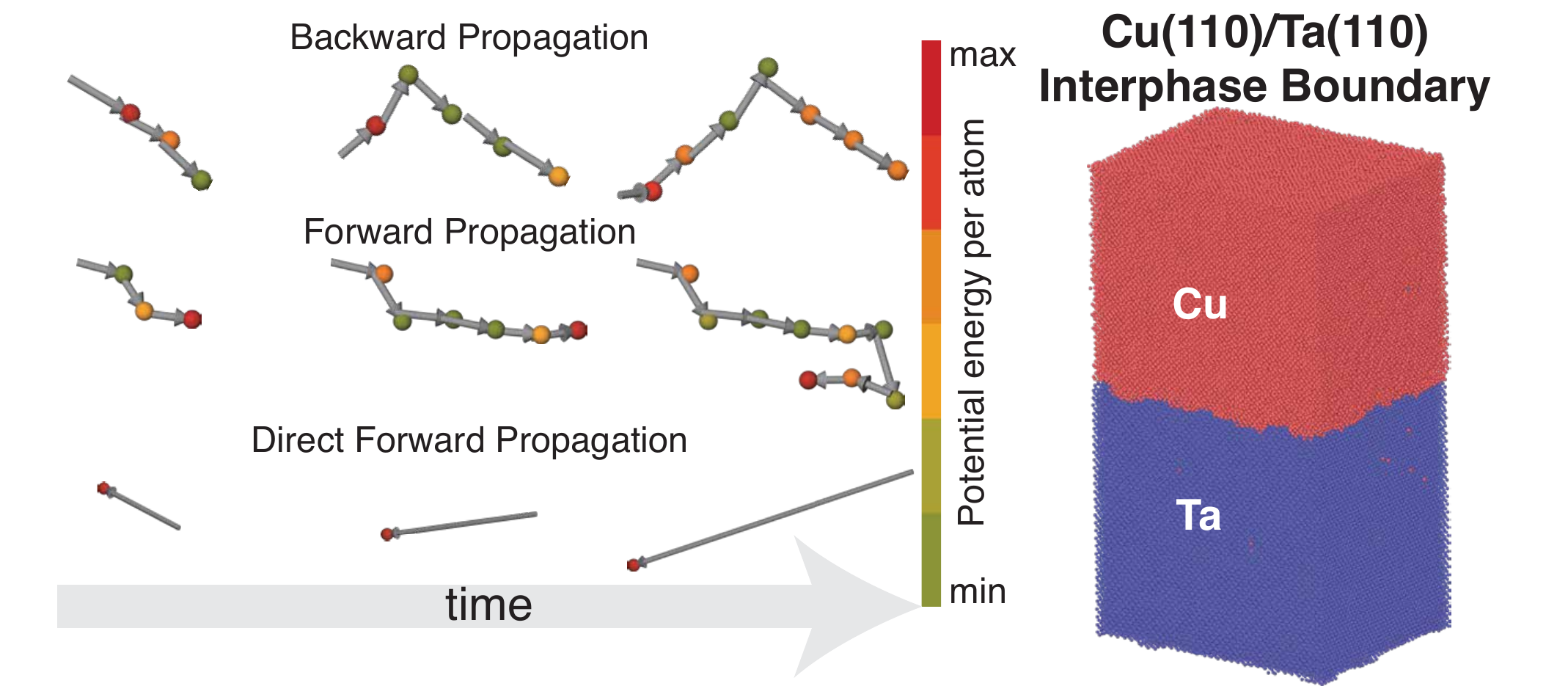}

\caption{String-like trajectories of atomic displacements at the Cu(110)/Ta(110)
interphase boundary at 1000 K. The model has the dimensions of $15.9\times12.6\times28.5$
nm$^{3}$. Periodic boundary conditions are applied along the directions
parallel to the interface.}
\end{figure}

\begin{figure}[H]
\begin{centering}
\includegraphics[width=0.8\textwidth]{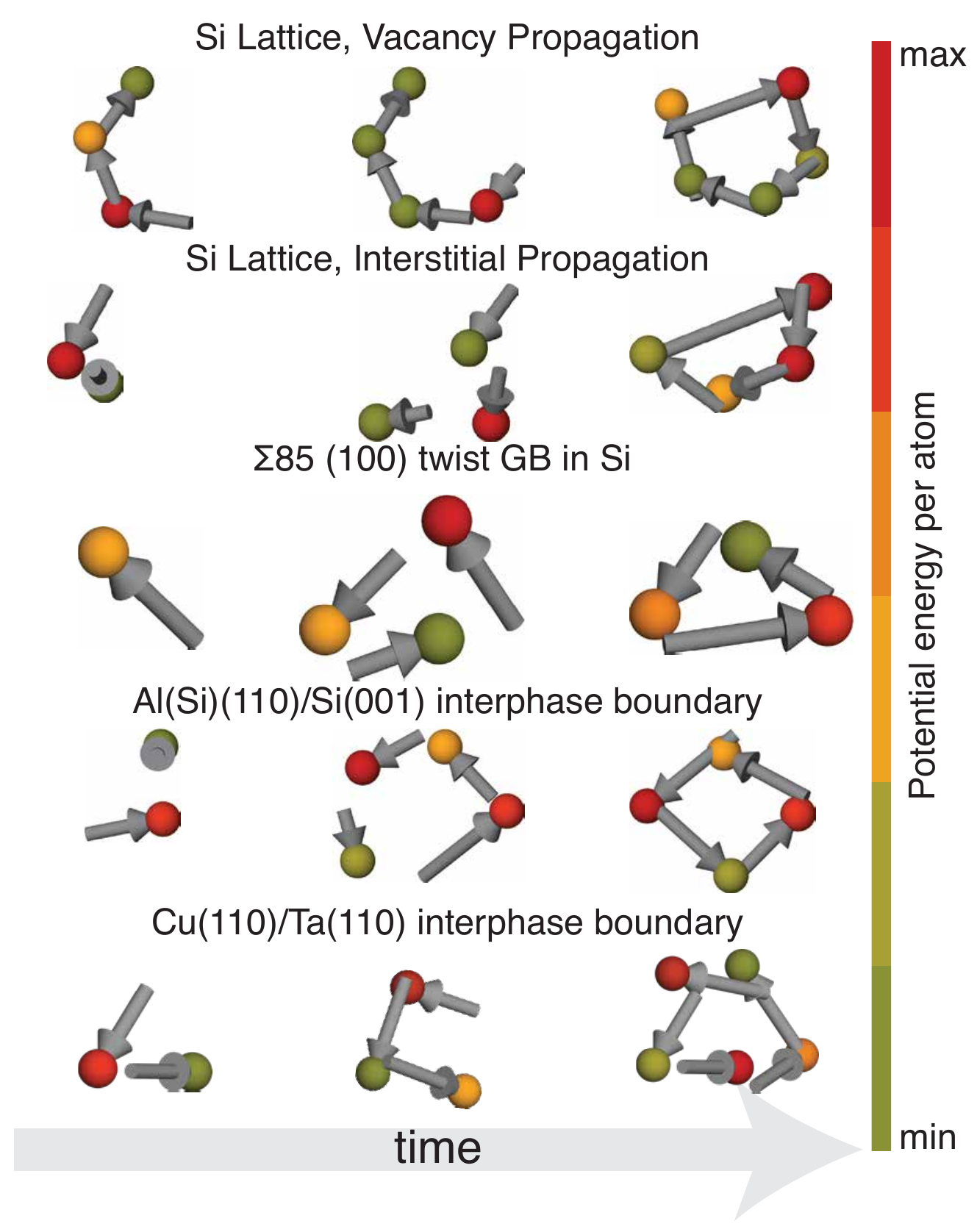}
\par\end{centering}
\caption{Closed string-like trajectories (rings) for the above-mentioned systems.}
\end{figure}

\end{document}